\documentclass[aps,prd,nofootinbib,preprint,tightenlines,showpacs]{revtex4}
\newcommand{\be}{\begin{equation}}
\newcommand{\ee}{\end{equation}}
\newcommand{\bea}{\begin{eqnarray}}
\newcommand{\eea}{\end{eqnarray}}
\usepackage{epsfig}

\begin{document}

\title{Is there any coherent measure for eternal inflation?}

\author{Ken D. Olum}
\affiliation{Institute of Cosmology, Department of Physics and Astronomy,\\ 
Tufts University, Medford, MA 02155, USA}

\begin{abstract}

An eternally inflating universe produces an infinite amount of spatial
volume, so every possible event happens an infinite number of times,
and it is impossible to define probabilities in terms of frequencies.
This problem is usually addressed by means of a measure, which
regulates the infinities and produces meaningful predictions.  I argue
that any measure should obey certain general axioms, but then give a
simple toy model in which one can prove that no measure obeying the
axioms exists.  In certain cases of eternal inflation there are
measures that obey the axioms, but all such measures appear to be
unacceptable for other reasons.  Thus the problem of defining sensible
probabilities in eternal inflation seems not be solved.

\end{abstract}


\maketitle

\section{Introduction}

Inflation is generically eternal.  While nearly every comoving
location in an inflating universe rapidly thermalizes, those that
do not are rewarded with continuing exponential expansion.  Except for
certain carefully arranged models, the physical volume of inflating
space grows forever, leading in turn to an infinite volume in
thermalized regions like ours.

Such a universe makes it difficult to make predictions.  There are an
infinite number of regions just like ours.  An infinite subset of
these will have any possible future.  So in what sense can we say that
one future outcome is more likely than another?

In a finite universe, even a very large one, anthropic ideas such as
Vilenkin's ``principle of mediocrity'' \cite{gr-qc/9406010} or
Bostrom's ``self-sampling assumption'' \cite{Bostrom:book} tell us to
consider ourselves a randomly chosen member of some finite \emph{reference
class} of observers.  But such an approach does not work with an
infinite class of indistinguishable regions.  There is no probability
distribution that assigns equal weight to each member of a countably
infinite set, so we cannot imagine ourselves to be drawn from such a
distribution.  Instead, we must use a \emph{measure}.

Many different measures have been suggested, but I will not attempt to
discuss the possibilities here.  Instead I will define the idea of a
measure by the job it has to do.  Suppose that we know the content of
the entire multiverse.  We also have a set of observations about
conditions here and in our past light cone.  Given these observations,
we can enumerate the places in the multiverse that we might be.
Unfortunately, there are infinitely many of them.  These different
places have various possible futures, say, $F_1, F_2,$\ldots.  A
measure is then a function that takes the entire multiverse and our
own current observations and returns a probability for each possible
future observation.

In principle, any such function is a measure.  But we want our measure
to have some degree of coherence, that is to say, its probabilities
should not be completely arbitrary.  The results of applying the
measure should have some correspondence to ordinary ideas of
probability.  A measure should not, for example, tell me one thing
based on my observations up to this moment and then without reason
change to something utterly different given my observations one second
from now.  In Sec.~\ref{sec:axioms}, I attempt to formalize this
idea by proposing two minimal axioms that I think any reasonable
candidate measure should obey.

Unfortunately, it turns out that these seemingly reasonable axioms are
not easily obeyed.  The difficulty is closely related to the problems
discussed in Refs.~\cite{Noorbala:2010zy,Bousso:2010yn,Guth:2011ie}.
In Sec.~\ref{sec:toy}, I describe a simple toy model of an infinite
universe and show that in that model no measure can satisfy both the
axioms.  In Sec.~\ref{sec:global}, I discuss more specifically why
global cutoff measures do not satisfy the axioms, unless one treats
the cutoff as a true ``end of time'' \cite{Bousso:2010yn} that renders
the universe finite.

In Sec.~\ref{sec:realistic}, I compare the toy model with more
realistic models of eternal inflation, and find that in certain cases
eternal inflation does admit measures that satisfy the axioms, but
these measures are undesirable for other reasons and fail if one
allows bubble collisions.  In Sec.~\ref{sec:Tiger}, I discuss the
consequences of weakening the requirement that the measure give an
observer the same probabilities after some time has passed without the
observer acquiring new information, but find that quite anomalous
situations can then arise.  I conclude in Sec.~\ref{sec:conclusion}.

\section{Axioms}\label{sec:axioms}

I now propose two axioms that I think any measure should satisfy in
order to have the resulting probabilities conform to our understanding
of how probabilities should behave.

\vspace{10pt}

\def\Bayesaxiom{1}
\def\consistencyaxiom{2}

\textbf{Axiom \Bayesaxiom a (Bayes's Rule --- general case)}
\emph{When we make new observations, the measure
should now yield the same probability distribution as we would get by
Bayesian updating of the old probability distribution using the new
observations.}
\vspace{10pt}

Specifically, suppose that the measure tells you that the probability
of some certain situation is $p(A)$.  Now we observe some process
whose outcome is determined (independently for each identical region)
by chance.  In case $A$, the chance to make some observation $O$ is
$p(O|A)$.  If not $A$, that chance is $p(O|-\!\!A)$.  If we observe
$O$, Bayes's rule will tell us that the chance to be in situation $A$
is now
\be\label{eqn:Bayes}
p(A|O) = \frac{p(O|A) p(A)}{p(O|A) p(A) + p(O|-\!\!A) p(-A)}
\ee
Applying the measure to our new state of knowledge should give this
same value for $p(A)$.

In fact, all we will need below is the trivial case of Bayesian
updating where you do not make any new observations:

\vspace{10pt}
\textbf{Axiom \Bayesaxiom b (Bayes's Rule --- trivial case)}
\emph{If time passes, but you receive no other new information, your
probabilities should remain unchanged.}
\vspace{10pt}

But could knowing that time has passed provide new information?
Indeed it could, if there were other alternatives.  For a mundane
example, suppose that you are about to have heart surgery.  The
surgeon tells you there is a 50\% chance that you have a minor problem
which will be easily repaired.  But there is also a 50\% chance that
you have a very severe problem.  The surgeon will try to repair it,
but your chance of surviving the surgery is quite small.  Now you're
anesthetized for the surgery.  If you wake up in the recovery room,
you will consider it very likely that you had only the minor problem.

In this case, $O$ is the observation that you are still alive.  The
observation $-O$, that you are dead, can never be made, but that does
not matter.  You should still update your probabilities according to
to Eq.~(\ref{eqn:Bayes}), with $A$ representing the minor problem and
$-A$ the severe one.

A more esoteric idea is that in certain circumstances you could be
duplicated.  Suppose that at 1:00 you flip a fair coin but do not
examine it.  You now think that the chance of heads is 1/2.  Now
suppose that at 1:10, if the coin is heads, there will be an exact
duplicate made of you (in a duplicate room, with a duplicate coin,
etc.)  If the coin is tails, nothing special will happen.

You wait until after 1:10.  Now I claim you should think the
probability of heads to be 2/3, since in the case of heads there are 2
copies of you sitting in rooms with coins showing heads, whereas in
the case of tails there is only one.  Again the passage of time has
given you new information, in this case about the possibility of
duplication\footnote{This problem is very similar to the ``sleeping
  beauty problem'', which has occasioned much philosophical debate.
  But it makes no difference whether you agree with my claim in this
  paragraph, so long as you accept the converse claim that if there is
  no possibility of duplication you shouldn't change your
  probabilities.}.

The important issue here is following.  In the first example,
your position is changed not simply by the passage of time but by the
possibility of a different outcome, in particular that your worldline
might have come to an end before the second observation.  The
second example is the reverse.  Your worldline might have come into
existence at 1:10.  In that case, you might also wish to readjust your
probabilities.  Thus we can update our axiom to exclude these
possibilities:

\vspace{10pt}
\textbf{Axiom \Bayesaxiom{} (Bayes's Rule --- final version)}
\emph{If the measure gives you a certain set
of probabilities at time $t_1$, you receive no new information
between $t_1$ and some later time $t_2$, and every worldline that
includes either time includes both, then the measure should give you
the same probabilities at $t_2$.}
\vspace{10pt}

``Receiving new information'' here means making an observation that
could have occurred differently (or not at all).  The passage of time
might be accompanied by other events that you knew would take place.
The fact that these events do take place as expected does not give you
new information.  Even though your circumstances are different after
seeing them, they do not constitute new information because you knew
that you would observe them, and thus they do not give a reason for
you to update your probabilities.

We will need one more axiom:

\vspace{10pt}
\textbf{Axiom \consistencyaxiom a (Consistency --- general case)}
\emph{If two or more observers are together and share their knowledge, they
should agree on the chance of future events.}
\vspace{10pt}

The idea here is that the measure is supposed to give probabilities of
future events.  If the event is to be observable by multiple
observers, the measure should not tell them to expect its occurrence
with different probabilities, unless some observers have relevant
knowledge that others lack.

In fact, we will need this only in the following more limited form.

\vspace{10pt}
\textbf{Axiom \consistencyaxiom{} (Consistency --- final version)}
\emph{If two or more indistinguishable observers are together, they
  should agree on the chance of mutually observable events.}
\vspace{10pt}

Since the observers are indistinguishable, the measure will
automatically supply the same probability distribution, but these
probabilities are relative to each observer.  For example, suppose two
observers know that exactly one of them has won the lottery.  Then the
measure will tell each observer that his own chance to be the winner
is the same number $p$, and the chance for the other person to have
won is $1-p$.  The consistency axiom then requires that $p=1-p$, so
$p=1/2$.

In addition to these axioms, the probabilities resulting from a
measure should obey the normal axioms of probability theory, i.e.,
they should be non-negative real numbers, the sum of the probabilities
of mutually exclusive possibilities should be the probability of their
union, and the total probability should be 1.

\section{Toy model}\label{sec:toy}

Now we ask whether there is any measure that satisfies axioms
\Bayesaxiom{} and \consistencyaxiom.  We begin by constructing a simple
toy model of an infinite universe.

Suppose there is a machine that produces observers.  At 1:00 it
produces 3 observers.  One of the 3 observers has a blue card in his
pocket, and other 2 observers have red cards.  Otherwise they are
identical.  Since they have not looked in their pockets, the 3
observers are subjectively indistinguishable.

Each observer is in his own bedroom, but the 3 (identical) bedrooms
are part of a single apartment with a common living room.  The doors
between the bedrooms and living room are open, so the observers can go
out to the living room and confer with each other.  All the observers
know all this information.

At 2:00, the machine produces another identical set of 3 observers,
and at 3:00 yet another set, and so on forever.  Thus the machine
creates a countably infinite set of subjectively indistinguishable
observers.  To keep the observers indistinguishable, they must not
know their creation time.  So we will endow each bedroom with a clock,
but these clocks will measure only the time elapsed for each observer.
Each clock will read 12:00 at the time that the observer is created.

Suppose you are one of these observers and you are wondering about the
color of your card.  You know that there are infinitely many observers
and that there are infinitely many with blue cards.  You cannot divide
these infinities to compute the chance that your card is blue.  You
need a measure.  The measure gives you some probability $p$ that your
card is blue.  Because the observers in the other bedrooms of your
apartment are subjectively indistinguishable from you, it must also
tell them that their cards are blue with the same probability $p$.
But because you and they can have a discussion, and because you know
the way in which you were all created, you can be certain that exactly
one of you is the ``blue observer'' (i.e., the one with a blue card).
Thus the measure must give you $p = 1/3$ to agree with the principles
of probability theory.\footnote{A previous version of this paper
  reached the same conclusion based on the symmetry of the creation
  process.  However, since the observers will later be treated
  differently based on their card color, it is not clear that such a
  principle can be used.  One might add as an axiom that the measure
  should not depend on future events, but this would rule out some
  measures currently under consideration, such as the light cone time
  or causal patch measure \cite{Bousso:2009dm}.}

Now we extend the scenario.  Some time after being created, each
observer goes back to his own bedroom, and the machine closes the
bedroom doors, separating the 3 observers in each apartment.  Then it
silently rearranges the bedrooms into two-bedroom apartments as follows.
For each creation step $N$, one of the red observers created by the
machine in that step will share an apartment with the blue
observer created in step $2N-1$, while the other red observer created in
step $N$ will share an apartment with the blue observer created in
step $2N$.  Thus every observer will be part of a pair, and every
pairing contains one blue observer.

When the clock in his room reads 12:30, each observer will see the
bedroom door open, and he will be able to talk to the occupant of the
one other bedroom in his apartment.  Since red observers must
generally wait to be paired with blue observers created many hours
later, time must run more slowly for them, so that they reach 12:30 at
the same time as their partners.  This could be implemented, for
example, via gravitational time delay.

All observers know all this, from the moment they are created.  Thus
each observer at 12:00 knows that he is now part of a triplet of
observers, and that when when his clock reads 12:30, he will be part
of a pair of observers.  He knows exactly how the pairings will be
made, but he does not know with whom he himself will be paired,
because he does not know which observer he is.

At 12:00, the measure told each observer that his chance to have a
blue card was 1/3.  What does it tell him at 12:30?  Let us check that
the conditions of axiom \Bayesaxiom{} are satisfied.  No observer
receives any information between 12:00 and 12:30.  At 12:30 the doors
open, but that has no bearing on the color of the cards.  It happens
at 12:30 independently of the colors, and the observer knew beforehand
that it would happen then.  Neither is there any worldline of any
observer which terminates or begins between times labeled 12:00 and
12:30.  Thus observers do not receive new information at 12:30, and so
by axiom \Bayesaxiom{} the measure must continue to tell each observer
that his chance to have a blue card is 1/3.

But this conflicts with axiom \consistencyaxiom.  Now that the doors
are again open, the observers can discuss their situations in pairs.
Each pair will conclude that exactly one of them has a blue card.
Each observer thinks that he has only a 1/3 chance to have a blue card
and so the other must have a 2/3 chance.  Consistency is violated.

Thus no measure can obey axioms \Bayesaxiom{} and \consistencyaxiom.  We
do not need to enumerate possible measures; we merely show what
results any measure must produce in order to obey the axioms, and show
that no measure can obey them all.

\section{Global time cutoff measures}\label{sec:global}

A common way to construct measures is to introduce a global time
coordinate $t$.  Only a finite number of observers exist at times before
some particular value of $t$, and thus probabilities can be defined as
frequencies if one considers only that subspace.  Then one takes
either a large value of $t$ or the limit as $t$ grows without bound
and one has a measure that gives well-defined probabilities.  

How would such measures handle our toy model?  First we have to
specify the choice of $t$.  Suppose first we use ``external time''.
We ignore the slow flow of time in some of the rooms and just consider
the time of creation, plus a fixed small amount of time after the
second observer in the pair is created, to bring the rooms together.
Thus an observer whose clock reads 12:00 considers herself typical
among all observers with such clocks that were created in the first $N$
steps of the machine.  One third of these observers have blue cards.

But the observer whose clock reads 12:30 reasons differently.  She
considers herself typical among those observers whose clocks have
reached 12:30 before some external time $N$.  Observers paired with
someone created after time $N$ are not included in this set.  Thus the
external time cutoff measure tells the observer whose clock reads 12:30
that her chance of having a blue card is 1/2.

The change from 1/3 to 1/2 violates axiom \Bayesaxiom.  The observer
has made no new observations, but she is nevertheless instructed to
change her probabilities.  A bizarre consequence of this sort of choice
will be discussed in section \ref{sec:Tiger}.

Now suppose instead we use a cutoff in proper time, by which we mean
the time before creation plus the subjective time spent in the room.
We consider all observers with proper time less than $N$ hours.  There
are $3N$ of these observers with clocks reading either 12:00 or 12:30.
So in either case, this measure yields probability 1/3.  The problem
here is that some observers who are in the reference class are able to
talk to observers who are not in the reference class.  This violates
axiom \consistencyaxiom.  Each observer thinks the following 3
possibilities are equally probable: ``My card is blue and yours is
red'', ``My card is red and yours is blue'', and ``My card is red and
your experiences don't count\footnote{There is some resemblance here
  to what philosophers call a ``zombie'', i.e., an entity who acts in
  every way as a person would, but somehow has no consciousness, so it
  is impossible to imagine that one is oneself that entity.}  because
you were created after time $N$''.  This nonsensical situation where
counted observers and non-counted observers can converse explains the
failure of the consistency axiom.

Bousso, Freivogel, Leichenauer, and Rosenhaus \cite{Bousso:2010yn}
argue that when we use a geometric cutoff, we must take it seriously
as an end of time.  In our case, this could be done by saying that
when the external time reaches, say, 10:59, the universe comes to an
end.  All observers' worldlines terminate there and the machine ceases
operation.  There are 30 observers created with clocks reading 12:00,
and 10 of those have blue cards.  Of these observers, 20, including
the 10 blue ones, are paired in apartments, while 10 of them terminate
at external time 10:59 without ever being paired.  Thus observers use
probability 1/3 at 12:00 and 1/2 at 12:30.  There is no contradiction,
because in this case the observers do learn new information at 12:30:
they learn that they have not reached the end of time.

This measure satisfies all the axioms.  That is possible because in
this case the universe is finite.  But for the same reason, no measure
is necessary.  Once the universe is finite, we can just consider
ourselves typical among all observers, without need for regularization.

Any finite universe will be free of measure problems, but I do not
think that constitutes a demonstration that our universe is finite.  I
argue here that we don't know what to do to construct sensible
probabilities in the case of an infinite universe, but the fact that
one does not know how to proceed in any given situation does not mean
that situation cannot exist.

Furthermore, even if the right answer is that the universe is finite,
I see no reason to introduce a cutoff intended to regulate infinite
universes and then argue that it is a physical boundary that renders
the universe finite.  One could equally well conclude that inflation
is not eternal or even that the universe will come to an end because
it is a giant simulation performed on a very large but finite
computer.

\section{More realistic models}\label{sec:realistic}

Could it be that eternal inflation is substantially different from
this toy model, so that an eternally inflating universe could be
handled by a measure, even though the toy model cannot?  One difference
is that the toy model generates universes linearly, whereas an
inflating universe grows exponentially.  But we can make a toy model
with exponential growth, as follows.

Instead of creating one blue and two red observers at each time, let
the machine create $2^{N-1}$ blue observers and $2^N$ red observers at
time $N$.  Each red observer created at $N$ can be paired with one of the
blue observers created at $N+1$.  There is still need for a time delay
in pairing the observers, but now it is merely that red observers need
to be delayed for 1 hour before being paired.  Otherwise the issues
are exactly as before and the measure cannot satisfy the axioms.

Another difference is that in eternal inflation observers are grouped
into different thermalized regions (``pocket universes''), which in
simple cases are each infinitely large.  The infinite number of
observers in each region can be handled by a measure which considers
only those in a very large spherical subregion.  As long as each
region is homogeneous, this part of the measure will pose no problem.

This technique handles the probabilities of the different observers in
the same region.  We still must handle the probability to be in one
region or another.  Techniques such as this, which treat the chance to
be a particular observer as the product of the chance to be in a
particular thermalized region and the chance to be one particular
observer within that region are called ``pocket-based measures''.
Several such ideas have been proposed
\cite{Vilenkin:1995yd,Easther:2005wi,Garriga:2005av}, but they have
mostly been abandoned because of several problems.

One such problem is that the pocket-based measures are prone to a
``$Q$ catastrophe'' in which the density perturbation amplitude $Q$ is
very strongly driven to either extremely large \cite{Feldstein:2005bm}
or extremely small \cite{Garriga:2005ee} values, giving a universe very
much unlike ours.

Another problem for these measures is that of Boltzmann brains
\cite{DeSimone:2008if}.  Each thermalized region begins with a big
bang that leads to the development of ordinary observers.  But
eventually each region with positive cosmological constant contains an
infinite, empty universe asymptotically approaching de Sitter space,
and an infinite number of Boltzmann brains can nucleate in any
comoving volume of that universe.  The reason pocket-based measures
avoid the problems discussed here (in particular violation of axiom
\Bayesaxiom) is that they never cut through the world line of any
observer, so they do not require observers to change their
probabilities at different times in their lives. But it is precisely
the property of not separating past from future in the same region
that leads to Boltzmann brain domination in the infinite future.

Finally, it is not really the case that within a thermalized region
conditions are the same everywhere.  Instead, the bubbles that
nucleate in an eternally inflating space-time generally collide with
each other and so break the equivalence between one part of the
thermalized bubble interior and another.

\section{Can we accept time-dependent probabilities?}\label{sec:Tiger}

Another possibility is that the axioms are wrong.  In particular,
Guth and Vanchurin \cite{Guth:2011ie} argue that one should simply
deny axiom \Bayesaxiom.  Instead, they explicitly allow situations
in which the probability one assigns to some state of affairs can
change merely because of the passage of time, even without any
possibility of death or duplication.

This idea leads to bizarre situations.  Once one accepts it, one can
construct a toy multiverse model which will lead to any desired
probability shift.  For example, let us consider the following thought
experiment.

It's exactly noon.  You are standing outside your house, which has two
doors.  You know that exactly one of these doors is locked.  Because
you are part of a multiverse of observers cooked up to produce
probability shifts, you believe at this moment that there is a 99\%
chance that the front door is the unlocked one.  But in 10 seconds you
will believe that the back door is the unlocked one with 99\%
probability.  Furthermore, you know that you will change your beliefs
at 12:00:10.  (Locked doors will remain locked and unlocked doors will
remain unlocked.  Your change of beliefs is entirely due to
re-evaluating your likely position in the multiverse as a measure
instructs you to do.)

Just at this moment a hungry tiger runs into your yard.  Your only
hope of survival is to get inside your house rapidly.  Naturally you
run for the front door, which you believe has a 99\% chance to be
unlocked.  On the other hand, if the tiger does not appear until
12:00:11, you run for the back door, which you now believe is the one
very likely to be unlocked.

But suppose instead that you are a 10-second run from either door when
the tiger appears at 12:00:01.  Because you have a head start, you'll
be able to reach either door a few seconds before the tiger.  Where
should you run?

Since when you see the tiger you think that the front door is
unlocked, perhaps you should run there.  But when you reach the
front door you'll be very unhappy because now you'll be nearly sure
it's locked.  Another possibility, recommended by Guth and Vanchurin
\cite{Guth:2011ie}\footnote{See their discussion of a subject who,
  before going sleep to for variable period of time, makes a bet that
  will be paid when he wakes up.  In fact the present example is
  strongly analogous to theirs, except in this case you must bet with
  your life.}, is that you should run for the back door, because you
know that by the time you arrive you will think it likely to be
unlocked.  But why, then, are you at 12:00:01 trying to escape a tiger
by running toward a door you're nearly sure is locked?  I think both
possibilities are nonsense, and therefore that one should not permit
measures that change probabilities only by the passage of time.

If the probability change were due to certain parts of the multiverse
ending, there would be no paradox.  The change of probability could
occur if there is a finite multiverse with a 99:1 preponderance of
regions with the front door unlocked, but nearly all those regions
terminate at 12:00:10, leaving a 1:99 distribution in favor of the
back door being unlocked.  In that case, you should think as follows.
``Most likely my worldline will end in 9 seconds, and this tiger does
not matter.  But in the small fraction of cases where my world line
continues, I'll be better off with the back door.''

There would also be no paradox if there were a timer scheduled to lock
and unlock doors at 12:00:10.  In that case your reasoning should be,
``The back door is probably now locked, but the timer will operate
before I reach the door, so the door will be unlocked when I reach
it.''

The situation with the multiverse measure is very different.  In that
case, you think the back door is likely locked, and you know that if
you run there you will arrive (not run into the end of time on the
way) and that the state of the door when you arrive will be the same
as it is now.  Thus when you set out for the back door, you feel
nearly certain of dying upon arrival.  Nevertheless, 1 second before
arriving, you will suddenly feel nearly sure of survival.  It is this
change, with no reason for it, that leads me to find such measures
unacceptable.

\section{Conclusion}\label{sec:conclusion}

To make sense of probabilities in an eternally inflating universe, we
need to introduce a measure, considered here as any procedure by which
probabilities can be assigned to an observer's circumstances.  I have
exhibited two simple axioms that I claim should be obeyed by any such
measure, and showed no measure can satisfy them in a toy model.  There
are some differences between eternal inflation and the toy model, but
it appears that these still do not allow us to construct any
acceptable eternal inflation measure.

What can we do to make sense of the situation?  One might try arguing
that anthropic ideas are wrong, and we should not consider ourselves
typical among all observers in a multiverse.  But then it does not
seem possible to recover ordinary ideas of probability.  In an infinite
universe, everything which can happen will happen an infinite number
of times, so what does it mean to say that one thing is more likely
than another?

Another idea is that of complementarity
\cite{Sekino:2009kv,Nomura:2011dt,Bousso:2011up}.  Perhaps it is not
meaningful to describe widely separated observers as simultaneously
having classical existence.  In the usual description, one could say
that there is an observer A at some point in the multiverse and an
indistinguishable observer B in some other place outside A's horizon,
and one needs a measure to decide one's chances of being A or B.
Complementarity claims that one can give a description of the universe
in which observer A exists, or a complementary description where B
exists, but one cannot describe them as existing at the same time, and
thus there is no need to decide which of these observers one is.
Complementarity does evade the difficulties I discuss here, but one
still needs some specific procedure to determine probabilities.

Finally, perhaps we just don't know how to proceed when the universe
is infinite (and not homogeneous).  I don't think one should conclude
from this that the universe must be finite, but rather that new ideas
are needed to make sense of probabilities in an infinite universe.

\section*{Acknowledgments}

I would like to thank Rafael Bousso, Ben Freivogel, Alan Guth, Vitaly
Vanchurin, Mike Salem, Ben Shlaer, and Alex Vilenkin for helpful
conversations.  This work was supported in part by the National
Science Foundation under grant number 0855447.

\bibliography{no-slac,paper}

\end{document}